\documentstyle[12pt,aaspp4]{article}
\begin{document}
\newcommand{\BB}{\mbox{${\bf B}_0$}}
\newcommand{\beq}{\begin{equation}}
\newcommand{\eeq}{\end{equation}}
\newcommand{\lsim}{\mbox{$\stackrel{<}{\scriptstyle\sim}$}}
\newcommand{\gsim}{\mbox{$\stackrel{>}{\scriptstyle\sim}$}}
\newcommand{\D}[2]{\makebox{$\displaystyle\frac{\partial{#1}}{\partial{#2}}$}}
\newcommand{\DD}[2]{\makebox{$\displaystyle\frac{\partial^2{#1}}{\partial{#2}^2}$}}
\input psbox
%\psonlyboxes
\title{Stochastic Acceleration of Low Energy Electrons
  in Plasmas with Finite Temperature.}
\author{Julia M. Pryadko\altaffilmark{1} and Vah\'{e} Petrosian\altaffilmark{2}}
\affil{Center for Space Science and Astrophysics\\
Stanford University\\
Stanford, CA 94305-4055}
\date{today}

\altaffiltext{1}{Department of Physics}
\altaffiltext{2}{Department of Physics and Applied Physics}

%\maketitle

\begin{abstract}
  This paper extends our earlier work on the acceleration of
  low-energy electrons by plasma turbulence to include the effects of
  finite temperature of the plasma.  We consider the resonant
  interaction of thermal electrons with the whole transverse branch of
  plasma waves propagating along the magnetic field.  We show that our
  earlier published results for acceleration of low-energy electrons
  can be applied to the case of finite temperature if a sufficient
  level of turbulence is present.  From comparison of the acceleration
  rate of the thermal particles with the decay rate of the waves with
  which they interact, we determine the required energy density of the waves as a fraction of the magnetic energy density,
  so that a substantial fraction of the background plasma electrons
  can be accelerated. The dependence of this value on the plasma parameter $\alpha = \omega_{pe}/\Omega_e$ (the ratio
  of electron plasma frequency to electron gyrofrequency), plasma
  temperature, and turbulence spectral parameters is quantified. 
  We show that the result is most sensitive to the plasma
  parameter $\alpha$.
  We estimate the required level
  of total turbulence by calculating the level of turbulence required
  for the initial acceleration of thermal electrons 
  and that required for further acceleration to
  higher energies.

\end{abstract}

\section{INTRODUCTION}

The importance of the stochastic acceleration of high energy charged
particles by turbulent plasma waves is well known. 
In our previous work (Pryadko and Petrosian, 1997, hereafter refered to as
PP1) we investigated 
the possibility of acceleration of the low energy background 
(thermal) electrons by this process using the well known formalism 
developed over the years and specifically the formalism described by 
Schlickeiser (1989) and Dung and Petrosian (1994). 
We considered interaction of electrons with plasma waves propagating
along the magnetic field lines in a cold plasma and presented analytic and 
numerical results on the acceleration and scattering time scales for 
different energies and plasma parameters.

The dynamics of the charged particles is determined by the
relationships between different plasma time scales which in turn are
governed by the properties of the background plasma. In a cold plasma
the most important time scales are the acceleration time $\tau_a$, the pitch
angle scattering time $\tau_{sc}$ and the time $\tau_{tr}\propto L/v$
needed for a particle with velocity $v$ to traverse the plasma region
of size $L$.  In cases when the pitch angle scattering time
$\tau_{sc}$ is much shorter than the traverse time $\tau_{tr}$ the
pitch angle distribution of the particles will be nearly isotropic.
The relative values of the acceleration and scattering times,
determined by the ratio of the pitch angle to momentum (or energy)
diffusion rates, also has important dynamic consequences. This ratio
varies strongly with the plasma parameters and the pitch angle and
energy of the electron. The solution of the transport equation when
this ratio is high is well known. In our previous paper (PP1) we
showed that this ratio becomes less than one at lower energies which
indicates that the usual transport equation derived for cosmic rays
may not be applicable for a wide range of energies and plasma
conditions.  A new transport equation was proposed for
non-relativistic electrons in high magnetic field, low density
plasmas.  In PP1 we also showed (numerically and through asymptotic
analytic expressions) that when this ratio is smaller than one the
acceleration time scale decreases with decreasing energy and could be
short; of the order of inverse gyrofrequency times the ratio of
turbulent to magnetic field energy densities.  The low energy
electrons are accelerated by their interaction with short wavelength
(high wave number) plasma waves.

In finite temperature plasma such waves 
will suffer a substantial damping (cyclotron damping).  Thus one more
time scale have to be taken into consideration - the damping time of
the waves.  If the damping rate of the waves by the background thermal plasma
is much less than the acceleration rate of the test particle, the
results obtained for the cold plasma with some small modifications
will be still applicable. In the opposite case the acceleration will
be limited to the decay time of the wave, so that for a sustained
process the wave production must occur at a faster rate than one would
consider for a cold plasma.

In this work we investigate the effects
due to the finite temperature of plasma. For this purpose we compare
the damping rate of the waves needed for acceleration of a low energy
electron with the acceleration rate of the same electron. We determine
the level of turbulent energy needed for acceleration of a substantial
fraction of the background plasma electrons under different plasma
conditions. In \S2 we reproduce some of the basic results obtained for
the cold plasma case. The modification of these results for finite
temperatures is discussed in \S3. In \S4 we evaluate the fraction of
the background electrons which are accelerated by the waves
for different plasma conditions. In \S5 we compare the levels of
turbulence needed for the initial acceleration of thermal electrons to
super-thermal energies with that required for further acceleration to
relativistic energies. A short summary is given in \S6.

\section{BASIC EQUATIONS FOR THE COLD PLASMA}

In a cold plasma 
at all energies the acceleration 
rate of the electrons is determined by the plasma parameters such as the
value of the magnetic field, the energy density of the plasma
turbulence, the spectral distribution of the waves and most importantly 
the plasma parameter $\alpha$, the 
ratio of plasma frequency $\omega_p$ to gyrofrequency of electrons $\Omega_e$:
\beq\label{alpha}
\alpha =\omega_{pe}/\Omega_e = \sqrt{\delta}/\beta_a = 3.2\ (n_e/10^{10}\ {\rm cm}^{-3})^{1/2}
(B /100\ {\rm G})^{-1}.
\eeq 
Here $\beta_a$ is the Alfv\'en
velocity expressed in units of speed of light and
$\delta = m_e/m_i$ is the ratio of electron to proton masses.

We are interested in acceleration of low energy electrons which are
interacting mainly with cyclotron and electromagnetic branch of the plasma
waves. The second interaction is possible only above some critical
energy and when $\alpha < 1$.  The electrons which do not have enough
energy to interact with the electromagnetic waves may still be in
resonance with whistler and/or electron-cyclotron modes. In what
follows we present the cold plasma relations for non-relativistic
electrons.

Following Dung and Petrosian, 1994 and PP1
and assuming a power law distribution of plasma turbulent energy density
as a function of wave vector,
${\cal E}(k)=(q-1){\cal E}_{tot}\ K_{min}^{q-1}\ K^{-q}$
(for $K \geq K_{min}$ and $q>1$), it can be shown that the acceleration 
time of the particle interacting with the plasma waves is:
\beq\label{tac}
\tau_a = \frac{p^2}{D_{pp}}=\tau_p \gamma^2 \frac{1}{(1-\mu^2)\sum_{j=1}^{N}
\left(\frac{\beta_{ph}(k_j)}{\beta}\right)^2\chi(k_j)},
\eeq
where
\beq
\chi(k_j)=\frac{|k_j|^{-q}}{|\beta\mu - \beta_{gr}(k_j)|},
\eeq
$\beta = v/c$ and the dimensionless wave vector $k_j$ 
is one of the roots (maximum of four) of the resonant condition:
\beq\label{res}
\omega(k_j)-\mu\beta k_j \mp 1/\gamma =0, \ \ \ \ k_j = K_jc/\Omega_e. 
\eeq
Here $\mu$ is the particle's pitch angle cosine, $\gamma$ is the
Lorentz factor and
the coefficient $D_{pp}$ is a momentum diffusion term that appears in the
well known Fokker-Planck equation.
The upper and lower signs refer to 
the right(R) and left(L) hand polarized plasma modes.
The wave frequency $\Omega$, in units of
electron gyrofrequency $\Omega_e$, 
is determined from the dispersion relation:
\beq\label{disp}
\frac{k^2}{\omega^2}= 
1-\frac{\alpha^2 }{(\omega \mp 1) \omega}, \ \ \
\  \omega(k_j)=\Omega(k_j)/\Omega_e,
\eeq
where, because for non-relativistic electron the resonant frequency
$\omega \gg \delta$, we have ignored the ion term of the order of  
$\delta$. The phase and group velocities of these
waves (in units of speed of light): $\beta_{ph}(k_j) = \omega_j/k_j$ and
$\beta_{gr}(k_j) = d \omega_j/d k_j$, respectively, can be
obtained from relations (\ref{res}) and (\ref{disp}). 

The parameter $\tau_p$,
which is a typical time scale in the turbulent plasma, is defined as 
\beq\label{taup}
\tau_p^{-1}=\frac{\pi}{2}\Omega_e\left(\frac{{\cal E}_{tot}}{B^2/(8\pi)}\right)
(q-1)k_{min}^{q-1}.
\eeq
The parameters important for our problem are $\Omega_e$, $\alpha$, $q, k_{min}$
and the ratio of plasma turbulent density to magnetic energy density:
\beq\label{fturb}
f_{turb}^{tot}=(8\pi{\cal E}_{tot}/B^2). 
\eeq
%The above equations hold for
%both electrons and protons. In what follows we consider only interaction
%and acceleration of electrons. 

Using the above equations it was shown  
in PP1 that for semi-relativistic electrons the resonant value of 
the wave vector is $k_{res} \simeq
(\alpha^2/\mu \beta)^{1/3}$ and the interacting wave propagates with phase
velocity $\beta_{ph} \simeq c/k$ and group velocity $\beta_{gr} \simeq 2
\mu \beta c$.
In this case 
the acceleration time of low-energy electrons with velocity $\beta$
interacting with 
electron-cyclotron and whistler waves, averaged over 
$\mu$, can be approximated by the analytic expression
\beq\label{acct} 
\langle\tau_a\rangle = \frac{2p^2}{\int_{-1}^{1} d \mu D_{pp}(\mu)} \simeq
\frac{(2+q)(8+q)}{18} \tau_p \alpha^{\frac{2(q+2)}{3}} \beta^{\frac{7-q}{3}}.
\eeq

\section{PLASMA WITH FINITE TEMPERATURE.}

Extrapolation of equation (\ref{acct})
 to zero energies will give zero
acceleration time for $q < 7$. However, such low energies are
in resonance manly with
electron-cyclotron waves with very high values of the wave vector. 
The cold-plasma dispersion relation allows the vector $k$ to become 
infinite at resonance, $\Omega \rightarrow \Omega_e$. However, in finite
temperature plasma this will not be possible because waves with 
such high values of $k$ are damped
quickly via cyclotron damping.  
Nevertheless, if the damping rate of the wave which accelerate
electrons of velocity $\beta$ and pitch angle $\mu$ 
 is less than the
acceleration rate of such an electron, the results obtained for the cold
plasma will be still applicable.

In this section we will investigate the range of plasma parameters for
which this condition is satisfied. We will be
interested in the case when electrons with energies as low as the
thermal energy of the background plasma can be accelerated.
To describe the thermal property of the plasma we introduce the
``thermal velocity'' in units of speed of light: 
$\beta_{th} = (kT/m_e c^2)^{1/2}$.

The dispersion relation for the electron cyclotron wave in plasma with
isotropic and finite temperature is
\beq\label{dist}
\frac{k^2}{\omega^2}=1-\frac{\alpha^2\delta}{\omega} +
\frac{\alpha^2}{\omega |k| \beta_{th}}\ Z(\zeta), \ \ \zeta=\frac{\omega
  -1}{|k| \beta_{th}},
\eeq 
where $Z(\zeta)$, the plasma dispersion function (see
e.g. Stix, 1992), can be expressed in terms of the complex error function:
\beq\label{z}
Z(\zeta) = i \sqrt{\pi} e^{-\zeta^2} (1+ Erf(i \zeta)).
\eeq
%\begin{eqnarray}\label{zzero}\makeatletter % This enables using @
%  Z_0(\zeta) = \left\{
%    \begin{array}[c]{lcr}
%           Z(\zeta),& k > 0\ , & \\
%           -Z(-\zeta) = Z(\zeta^*)^*,&  k < 0. &
%    \end{array}\right. % This removes the equation number
%\end{eqnarray}
 The second term on the right hand side of
(\ref{dist}) describes the ion contribution and is negligible for 
waves with $\omega \simeq 1$.
Since electron-wave interaction is described by the resonant
condition (\ref{res}) we can determine the decay rate of the wave in
terms of the energy of the interacting electron.

It is well known that for high energy electrons, $\beta \gg\beta_{th}$,
and for typical pitch angles, $\mu \neq 0$, the parameter $\zeta \gg 1$ and
the imaginary part of the dispersion function is negligible (see
e.g. Gershman, 1953 or Stix, 1958). 
In this case the dispersion relation
(\ref{dist}) reduces to the cold plasma dispersion relation
(\ref{disp}). Therefore, for electrons with kinetic energies much greater than the
thermal energy the results obtained with the assumption of a cold plasma
are valid. For particles with velocities close to the thermal velocity we
have to compare their acceleration time with the decay time of the
wave that they interact with to make sure that there is enough time for the
particle to get accelerated before the wave disappears.

The usual technique to calculate the wave decay rate is to separate
the frequency $\omega$ into real and imaginary parts,
$\omega = \omega_r + i \omega_i$, and use it in equation (\ref{distt}).

\subsection{Super-thermal non-relativistic electrons}

First we consider the case when $| \omega_i | \ll | \omega_r -1 |$,
which as we show below is valid for particles with velocities several
times the thermal velocities.
In this case the imaginary part of the parameter $\zeta$ is negligible
so that, to the second order in $\beta$, substitution of the resonant
condition in equation (\ref{dist}) gives $\zeta = \mu \beta/\beta_{th}$,
independent of $\omega_r$, $\omega_i$ or $k$. The
dispersion function $Z(\zeta)$ then can be separated into real and imaginary
parts:
\begin{eqnarray}\label{distt}
Z(\zeta)& = &\frac{X + iY}{\zeta}, \\
\ X\equiv -2\zeta S(\zeta),\ \ Y&\equiv&\sqrt{\pi}\ \zeta \ e^{-\zeta^2},
\ \ S(\zeta)\equiv e^{-\zeta^2}\  \int_0^\zeta dz
e^{z^2}. \nonumber
\end{eqnarray}
In the limit under consideration here both functions $X$ and $Y$ are independent
of $\omega_r$ and $k$.
Substituting $\zeta = \mu \beta/\beta_{th}$ into the above equations and
neglecting the ion contributions (order of $\delta \ll \omega_r$) we obtain:

\begin{equation}\label{omegar}
%\makeatletter % This enables using @ 
%    \begin{array}[c]{lcl}
      \frac{k^2}{\omega_r^2}=1-\frac{\alpha^2}{\omega_r (\omega_r-1)}\left\{X
      -\frac{\alpha^2 Y^2}{k^2-\alpha^2 X}\right\}
                                % This puts a label (XXXa)
%     \hbox to0pt{\hskip1.9in(\theequation a)\hss} \\
\eeq
\beq\label{gamma}
     \omega_i=\frac{\alpha^2 Y}{k^2 - \alpha^2 X}\omega_r.
                                % This puts a label (XXXb)
%      \hbox to0pt{\hskip2.67in(\theequation b)\hss}
%    \end{array} % This removes the equation number
%  \def\@eqnnum{\relax}
\end{equation}
Equation (\ref{omegar}) describes the modified dispersion
relation for the real part of $\omega$ and the equation (\ref{gamma}) gives
the decay rate of the waves.  We can solve equation (\ref{omegar}) for
$k$ and substitute it into equation (\ref{gamma}) to obtain the
damping rate.  We are interested in particles with energies close to the
thermal energy, $\beta \sim \beta_{th}$, so that the argument
$\zeta$ and the functions $X$ and $Y$ are of order of unity and $k^2
\simeq (1-\omega_r)^{-1} \gg 1$.  Taking this into consideration
the dispersion relation simplifies to 
\beq\label{dissim}
\frac{k^2}{\omega_r}\simeq\frac{\alpha^2}{\omega_r-1} X. 
\eeq 
The
difference between this dispersion relation and the cold plasma
dispersion relation (\ref{disp}),
in the limit $\omega_r \rightarrow 1$, is that the
term $\alpha^2$ is replaced by $\alpha^2X$.  Since $X$ is independent of
$k$ this fact allows us to use the value of the resonant vector
obtained for the cold plasma case, $k_{res}^3\simeq \alpha^2/(\mu \beta)$, as a
solution in equation (\ref{dist}). This is achieved by simply replacing $\alpha^2$ by
$\alpha^2 X$ in equations (\ref{omegar}) and (\ref{gamma}) which gives
\begin{equation}\label{krest}
%\makeatletter % This enables using @ 
%    \begin{array}[c]{lcl}
      k_{res}^3\simeq\frac{\alpha^2 X}{\mu\beta},
                                % This puts a label (XXXa)
%      \hbox to0pt{\hskip2.55in(\theequation a)\hss} 
\eeq
\beq\label{gamrest}
      \omega_i\simeq\frac{(\mu \alpha
        \beta)^{\frac{2}{3}}Y}{X^{\frac{2}{3}}}\  \omega_r
                                % This puts a label (XXXb)
%      \hbox to0pt{\hskip2.35in(\theequation b)\hss}
%    \end{array} % This removes the equation number
%  \def\@eqnnum{\relax}
\end{equation}
Using the above equations we can now evaluate the ratio of the
imaginary to real part of $\zeta$: $| \omega_i |/| \omega_r -1 | =
Y(\zeta)/X(\zeta)$.  To be consistent with the assumption stated above
this ratio has to be much less than unity. For $\zeta = \mu
\beta/\beta_{th} \simeq 0.8 $ we get $Y = X$, therefor the above
solution is restricted to velocities greater than several times the
thermal velocity (depending on the pitch angle).

In order to compare the wave decay time and the
acceleration time we will average the decay rate $\omega_i$ over the
pitch angle to get an averaged decay time $\tau_{dec}\equiv
1/<\omega_i>$.  Introducing the ratio of the acceleration time to the
decay time of the wave, $R =\tau_a/\tau_{dec}$, and using equations
(\ref{acct}) and (\ref{gamrest}) we obtain an approximate dependence of
$R$ on the plasma parameters and temperature;
\beq\label{anrat} 
R \simeq
\alpha^{2+\frac{2q}{3}}\beta_{th}^{3-\frac{q}{3}} \Omega_e
  \tau_p.  
\eeq 
For
successful acceleration we need $R$ to be less than one or,
equivalently $\tau_p$ to be less than some value
which will depend on temperature, plasma parameter $\alpha$ and the spectral
index $q$.
Since $\tau_p^{-1}$ is proportional to the fraction of plasma
turbulence $f_{turb}^{tot}={\cal E}_{tot}/(B_0^2/8\pi)$, this requirement sets a lower
limit on the amount of turbulent energy that is needed for
acceleration of plasma electrons with energies of several times the
thermal energy.

\subsection{Thermal electrons}

In order to obtain a solution for the acceleration time of electrons 
with $\beta \simeq \beta_{th}$ when the
condition $|\omega_i| \ll |\omega_r -1|$ is not valid we have to use a
 complex $\zeta$ in equation (\ref{distt}):
\beq\label{zeta}
\zeta = \zeta_r + i \zeta_i, \ \ \ \zeta_r =\frac{\omega_r -1}{k\beta_{th}}=\mu \frac{\beta}{\beta_{th}}, \ \ \ \zeta_i=\frac{\omega_i}{k\beta_{th}}. 
\eeq 
The separation of real and imaginary parts of the dispersion relation in 
this case results in two equations which are cubic in $k$ and
nonpolinomial in $\omega_i$ so that
the solution for $k$ and $\omega_i$ can only be found numerically.
The resultant solutions show that as in the cold plasma electrons with
velocities of order of thermal velocity interact with waves of frequency close to
the electron gyrofrequency,  but the resonant
wave vector now is of order of unity as opposed to the $k_{res} \gg 1$
values for the zero temperature case (see Figure \ref{kres}).
Using the new dispersion relations we then calculate the phase and
group velocity of the resonant waves. Substituting these in
equation (\ref{tac}) gives
the acceleration time assuming that 
the power law ${\cal E}(k) \propto k^{-q}$ is still applicable. 
Equating the acceleration time with the decay time of the waves
($1/\omega_i$) gives a general expression for the critical values of
several parameters.

Clearly, given an initial distribution of the wave vector $k$
(e.g. power law) the damping and acceleration will modify the wave
spectrum in time and space. An exact solution of the problem 
requires treatment of
the coupled wave-particle kinetic equation. This is beyond the scope
of the present paper and will be considered in future publications. 
The assumption of a power law spectrum will suffice for our purpose
here. As we shall see below the results turn out to be almost
insensitive to the spectrum.

%Since the simple analytical solution is not possible in this case and is
%very computationally intensive we will instead find the level of plasma 
%turbulence at which the acceleration time is much smaller than the inverse of
%the damping rate of the waves so that we can use the approximation 
%(\ref{acct}) found for the cold plasma. 

\section{ACCELERATED FRACTION OF THE BACKGROUND ELECTRONS}

%In this section we assume that a power law wave spectrum is maintained
%up to a critical wave vector $k_{cr}$ (at which the damping and
%acceleration rates are equal). Beyond this we expect the wave spectrum
%to drop off abruptly (which is a reasonable approximation since the
%waves with the higher wave numbers are strongly damped). The energy
%and pitch angle range of electrons which are in resonance with waves
%of $k < k_{cr}$ then determine the fraction of electrons that are
%accelerated efficiently.  

Our goal is to determine the plasma conditions at which the number of
the accelerated electrons is a significant fraction (say, greater than
10\%) of the background electrons. In particular, we want to determine
the required level of turbulence for acceleration of such a fraction.
  From the condition for efficient
acceleration, $\tau_{a} \omega_i \leq 1$, we can find an upper limit
on the characteristic time scale $\tau_p$ which in turn gives the
minimum required plasma turbulence energy density
$f_{turb}(\mu,\beta,\alpha,T,q)$.  The relation between $f_{turb}$ and
$\tau_p$ depends on the spectral distribution of the waves. For a
power law spectrum this relation is given by equation (\ref{taup}).
Note that there is an additional dependence of $f_{turb}$ on the wave
spectrum due to the dependence on the spectrum of the normalized acceleration
time $\tau_a/\tau_p$. As we will show later this effect is small;
$f_{turb}$ decreases slightly with increasing $q$. The main variation
of $f_{turb}$ comes from the value of $k_{min}$ in equation
(\ref{taup}). The criterion $\tau_a \omega_i \leq 1$ then translates
to $f_{turb}^{min}(k > k_{min}) \propto k_{min}^{1-q}$. The value of
$k_{min}$ is unknown but it should be at least as low as the minimum
wave number of waves in resonance with electrons in the
considered range of energies. 
In this section we concentrate on the dependence of $f_{turb}^{min}$
on other plasma parameters and assume $k_{min} = 1$. Later we will
show how to modify the results using the correct value of $k_{min}$.
So in what follows $f_{turb}$ represents 
$(8 \pi {\cal E}_{turb}/B^2) k_{min}^{q-1}$.
The total turbulent energy, due to all the branches
including those with $k < k_{min}$ will be, of course,  
larger than the above estimate. We shall return to this in the
next section.

\begin{figure}[h]
\leavevmode\centering
\psboxto(\textwidth;0pt){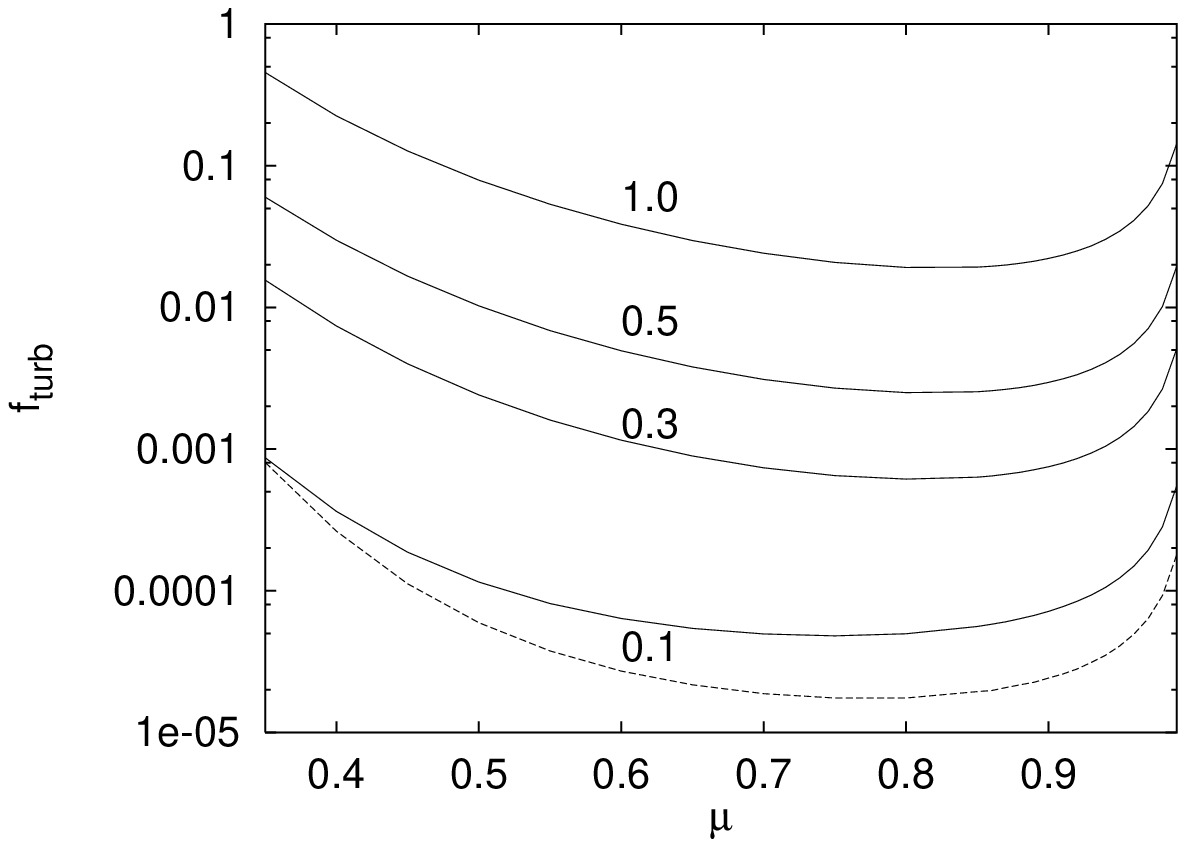}
\caption{The ratio of plasma turbulent energy density, 
to magnetic energy density $f_{turb}=(8\pi{\cal E}_{tot}/B^2)
k_{min}^{q-1}$, at
which the wave damping rate is equal to the electron acceleration rate,
as a function of $\mu$ for plasma parameter $\alpha = 1$, $0.5$, $0.3$
and $0.1$ from top to bottom.
The plasma temperature is assumed to be 1 keV and $\beta =\beta_{th}$.
The spectral index of the wave turbulence is $q=5/3$ except for the
dashed line for which $q=3$ and $\alpha = 0.1$.}
\label{ratA}
\end{figure}

\begin{figure}[h]
\leavevmode\centering
\psboxto(\textwidth;0pt){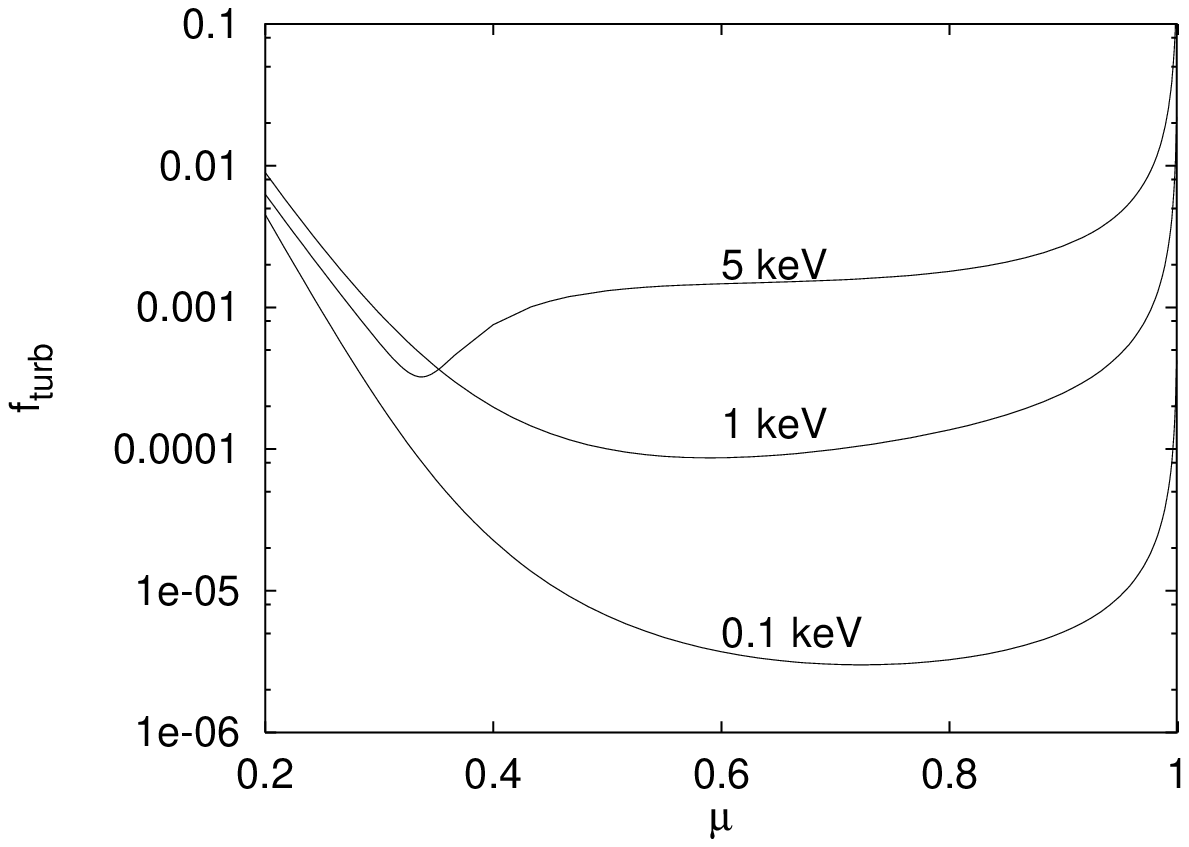}
\caption{Same as Figure \protect\ref{ratA} except for the different values of
  temperature and for $\alpha=0.1$, $\beta = \beta_{th}$, $q=5/3$.}
\label{ratT}
\end{figure}

Following the above procedure we calculate the variation of $f_{turb}$
with $\mu$ for a thermal
electron ($\beta = \beta_{th}$) and different values of $\alpha$ and
for several values of temperature. The results are shown on Figures \ref{ratA} and
\ref{ratT}.  It can be seen that in order to have an acceleration time
at least as short as the decay time of the waves the level of the
turbulence has to be greater than some threshold value
$f_{turb}^{*}$, corresponding to a minimum of the curves at some
value $\mu_{*}$ of the pitch angle cosine.  If the level of turbulence
is less than this value then the acceleration process will
not be efficient because the interacting waves are decaying too fast.
However, because the curves in Figures \ref{ratA} and \ref{ratT} are
nearly flat over a wide range of $\mu$, a slightest increase in the
turbulent level above this threshold value will result in
accelerating electrons with a wide range of pitch angles.

\begin{figure}[h]
\leavevmode\centering
\psboxto(\textwidth;0pt){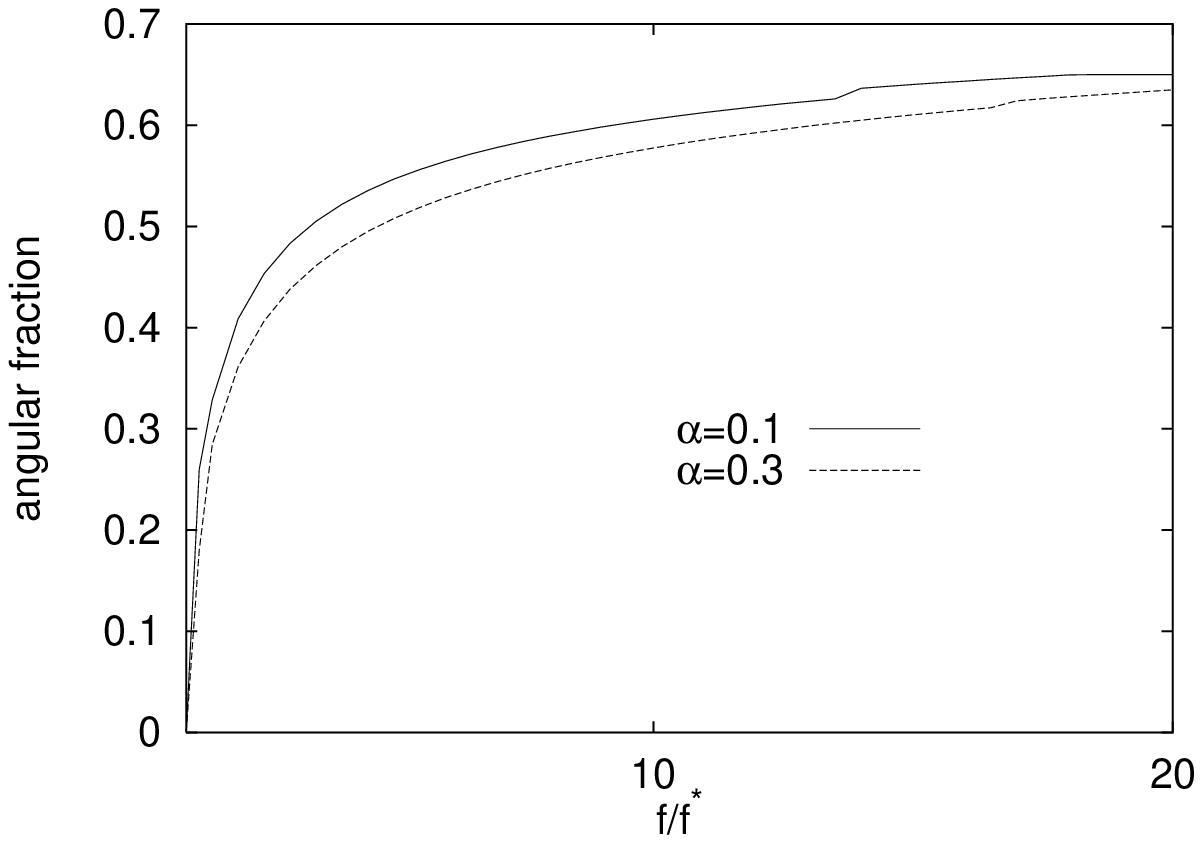}
\caption{The fraction of pitch angle range of isotropicly distributed electrons with $\beta/\beta_{th}=1$ 
involved in the acceleration process as a function of the ratio of the
turbulent level $f$ to the minimum turbulent level $f_{*}$ (obtained
from Figures \protect\ref{ratA} and \protect\ref{ratT}) 
in plasma with $T=1$ keV.}
\label{frac}
\end{figure}

%This is demonstrated in Figure \ref{frac} which shows 
The fraction  of the electrons which are accelerated
more rapidly than the waves decay is proportional to $0.5
(\mu_2 -\mu_1)$ where $\mu_2$ and $\mu_1$ are the high and low values
of the intersection with the curves on Figures \ref{ratA} and
\ref{ratT} of a horizontal line corresponding to a given value of the
$f_{turb}$. The variation of this fraction with $f_{turb}$ is shown
in Figure \ref{frac} for $\beta=\beta_{th}$ at $T \simeq 1$ keV.
As evident this fraction increases rapidly with
increasing $f_{turb}$ up to a level about 2.5 times the minimum level,
and then changes slowly. Thus a level of the wave turbulence equal to
$3 f_{turb}^{*}$ will involve more than half of the electrons in
the acceleration process. For an isotropic distribution the vertical
axis of Figure \ref{frac} is proportional to the fraction of electrons
accelerated, and for a Maxwellian distribution the proportionality constant for $\beta=\beta_{th}$ is
$(4/\sqrt{\pi}) \int_1^{\infty} e^{-x^2} x^2 dx$.

% As the fraction of
%the turbulent energy increases the range of the pitch angles of
%the accelerated electrons becomes wider for every given particle's energy.

%To show the sensitivity of the fraction of electrons which have
%acceleration time no less than the wave decay time to the
%ratio $f/f_{min}$ we plot on Figure
%\ref{frac} the fraction of an isotropic
%distribution of electrons with energy $kT$ ($r=1$).  

\begin{figure}[h]
\leavevmode\centering
\psboxto(\textwidth;0pt){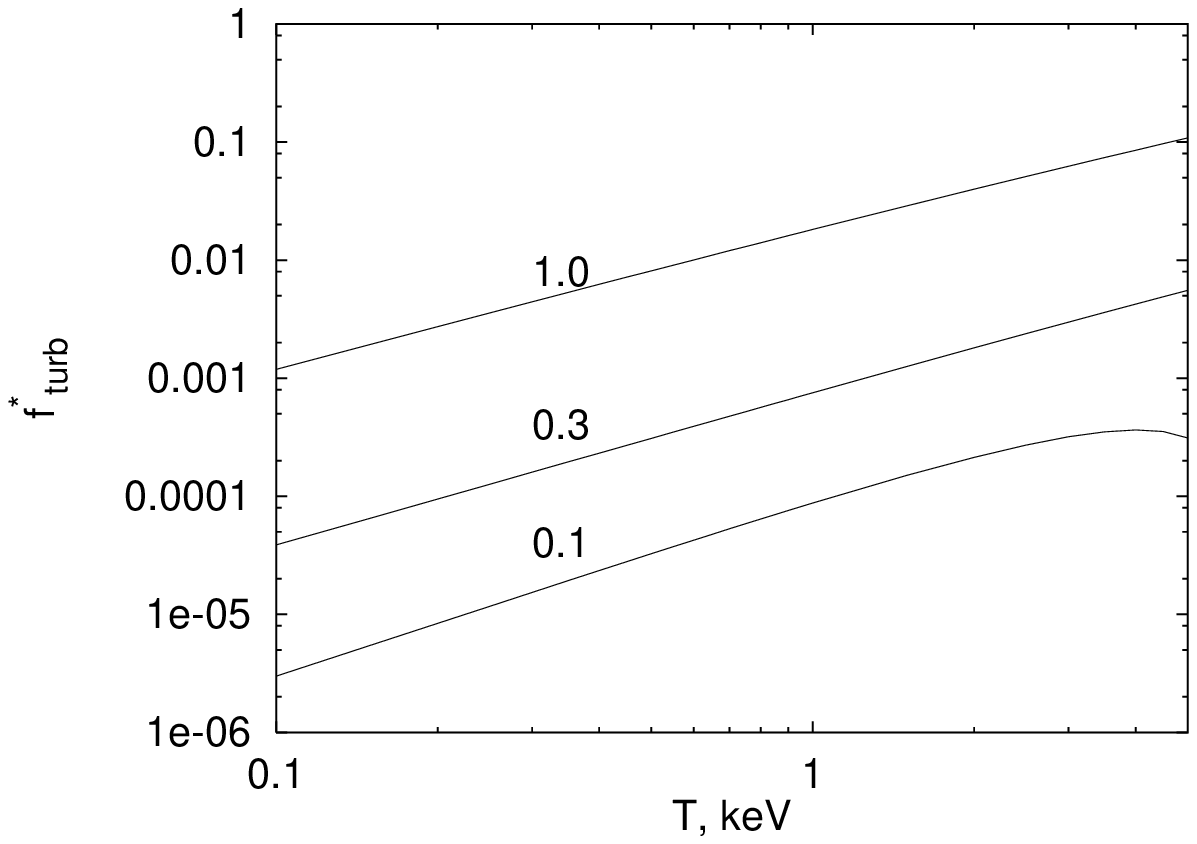}
\caption{The dependence of the minimum of turbulence fraction $f^{*}_{turb}$ 
on plasma temperature for electrons with $\beta=\beta_{th}$
in plasma with $q=5/3$ and three different values of the parameter $\alpha$.}
\label{tmin}
\end{figure}

\begin{figure}[h]
\leavevmode\centering
\psboxto(\textwidth;0pt){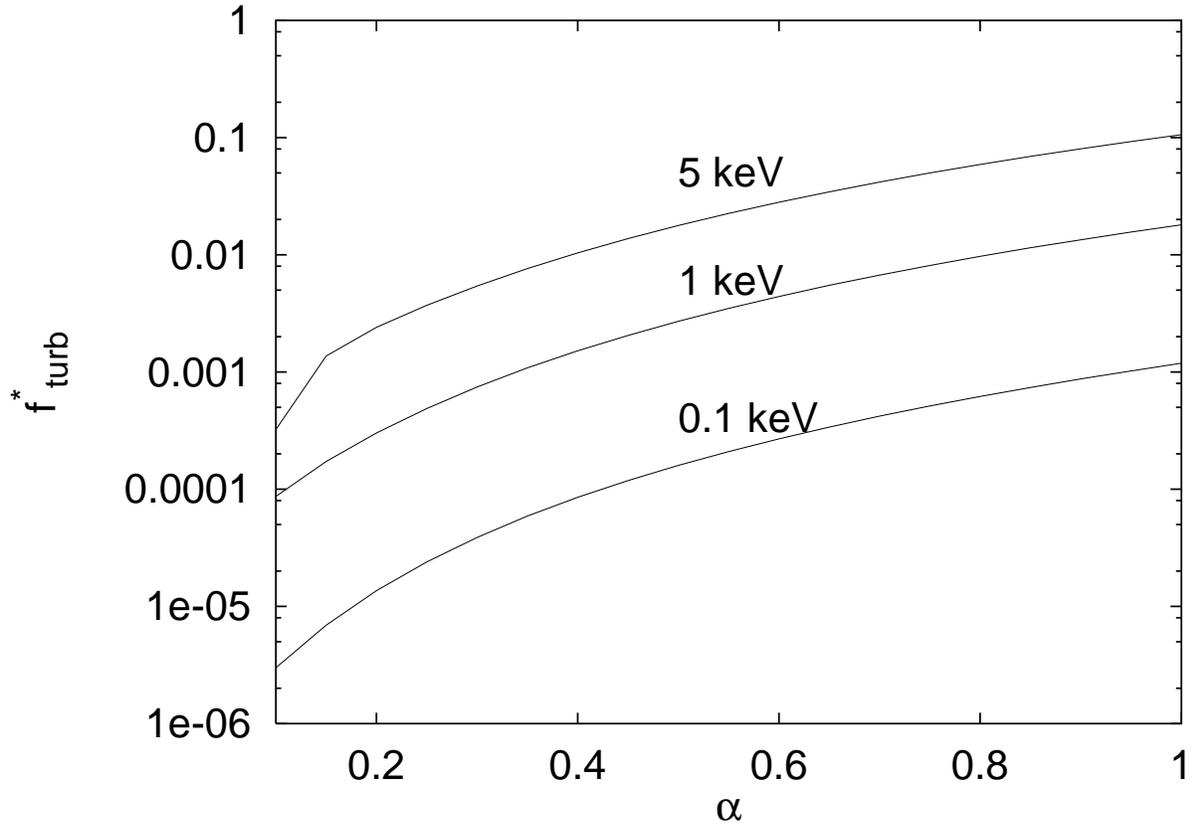}
\caption{The dependence of the minimum of turbulence fraction $f^{*}_{turb}$ 
on plasma parameter $\alpha$ for $q=5/3$, $\beta=\beta_{th}$ and three
different temperatures.}
\label{amin}
\end{figure}

\begin{figure}[h]
\leavevmode\centering
\psboxto(\textwidth;0pt){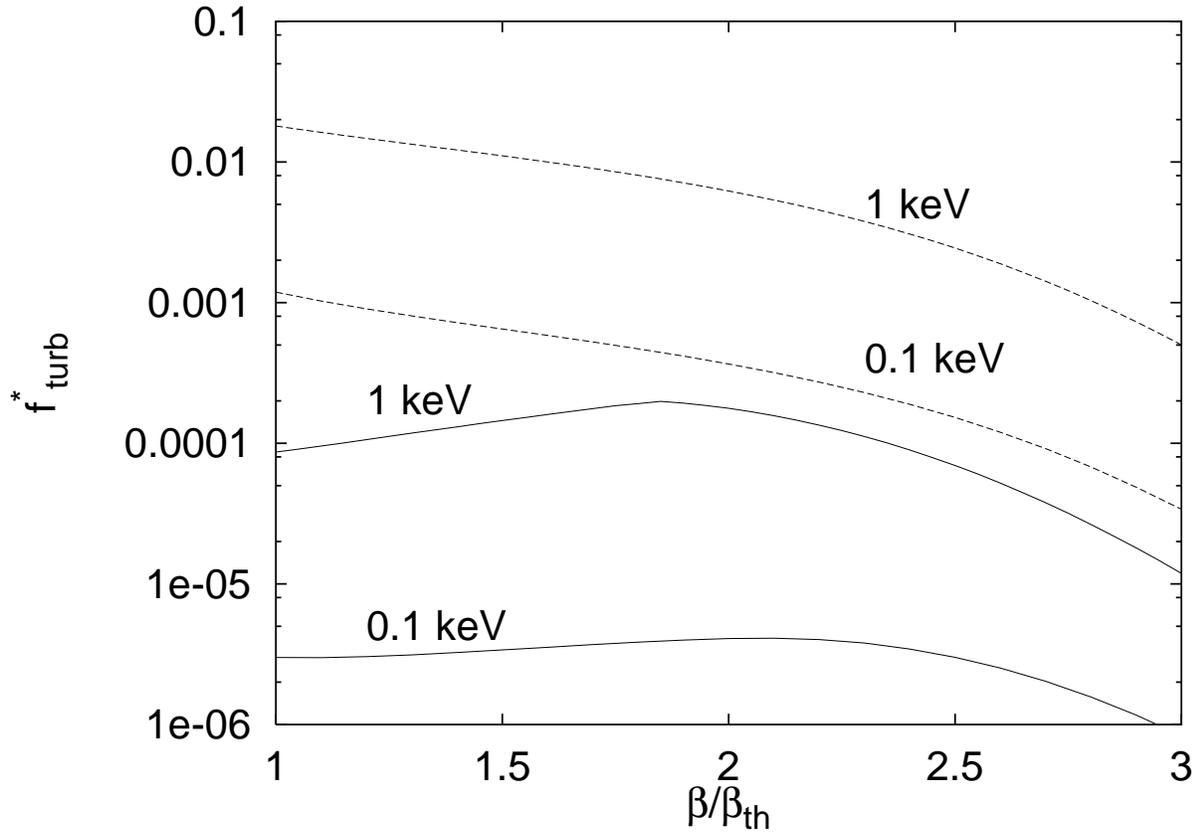}
\caption{The dependence of the minimum of turbulence fraction $f^{*}_{turb}$ 
on electron velocity (in units of thermal velocity) for $q=5/3$ and different
values of plasma parameter and temperature. Solid lines are for
$\alpha = 0.1$ and dashed lines are for $\alpha=1$.}
\label{rmin}
\end{figure}

\begin{figure}[h]
\leavevmode\centering
\psboxto(\textwidth;0pt){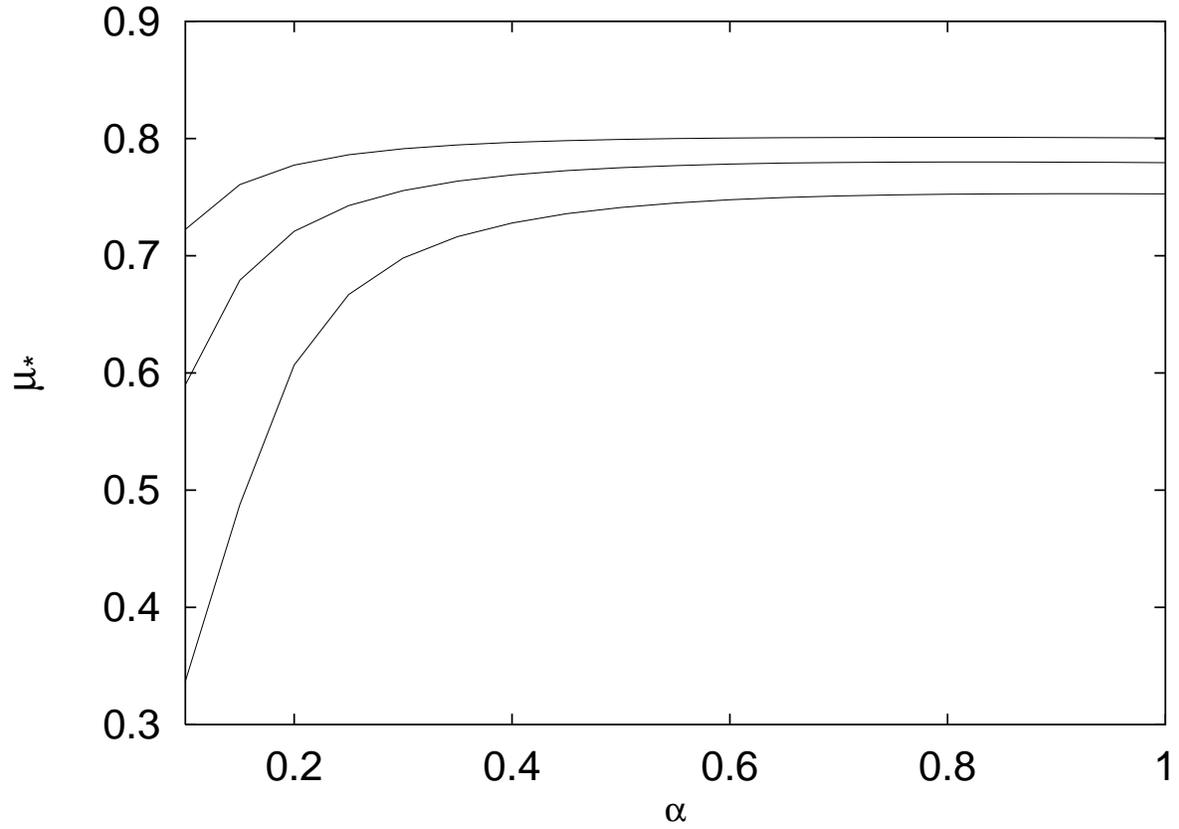}
\caption{The dependence on the plasma parameter $\alpha$ 
of the critical pitch angle, at which the turbulent fraction
$f_{turb}$ is minimum, for $q=5/3$ and three different 
temperatures, $0.1$, $1$ and $5$ keV, from top to bottom.}
\label{mumin}
\end{figure}

\begin{figure}[h]
\leavevmode\centering
\psboxto(\textwidth;0pt){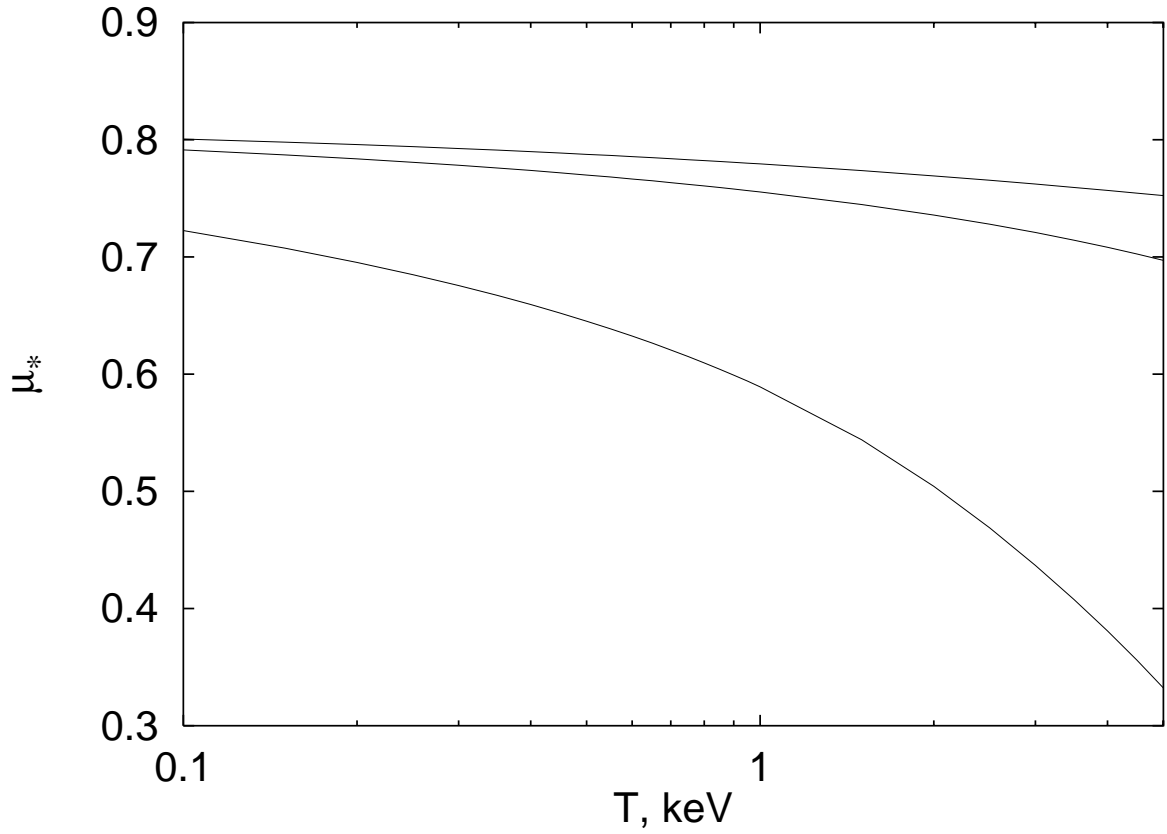}
\caption{Same as Figure \protect\ref{mumin} but $\mu_*$ versus plasma
  temperature for $\alpha = 1$, $0.3$, $0.1$ from top to bottom.}
\label{muminT}
\end{figure}

\begin{figure}[h]
\leavevmode\centering
\psboxto(\textwidth;0pt){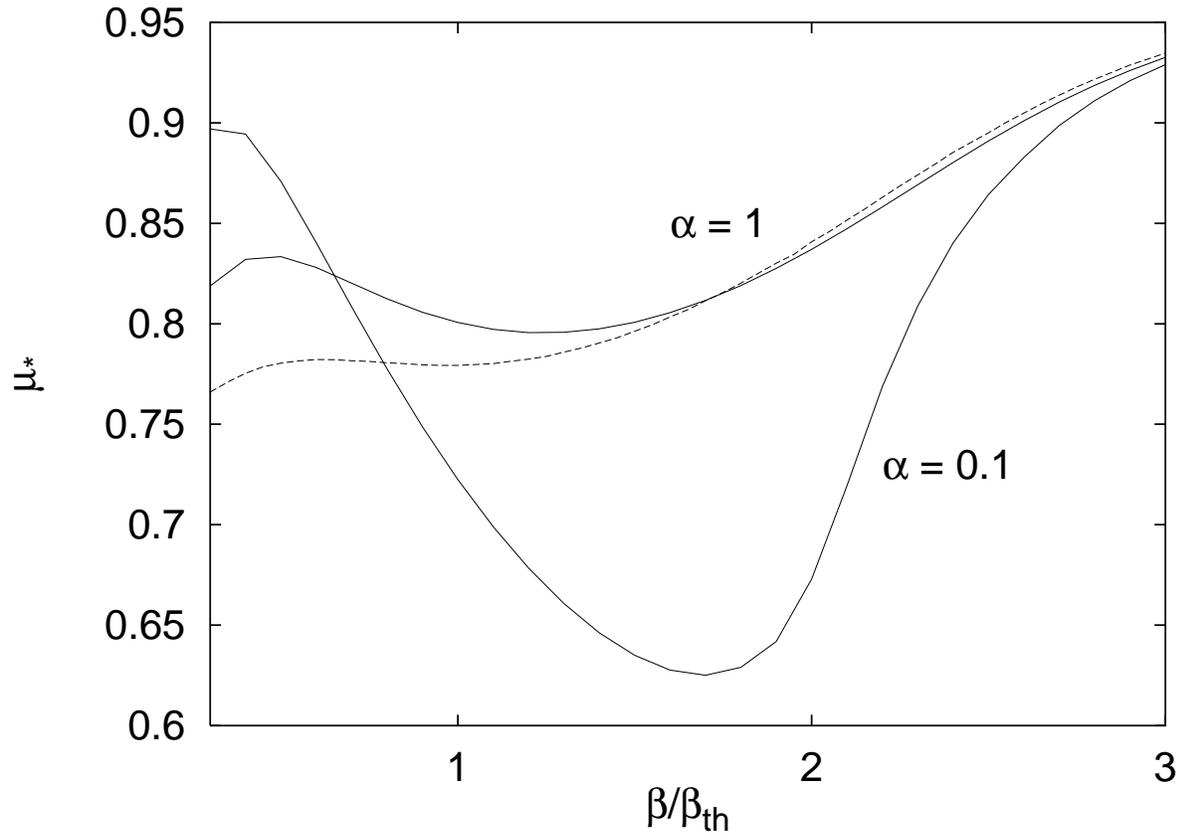}
\caption{Same as Figure \protect\ref{mumin} but $\mu_*$ versus
  electron velocity for indicated values of $\alpha$ and $T = 0.1$ keV
  (solid lines), $T = 1$ keV (dashed line).}
\label{muminR}
\end{figure}

It is clear, therefore, that knowing the value of $f_{turb}^{*}$ will
allow us
to estimate the effectiveness of the stochastic acceleration. Consequently, on Figures \ref{tmin} and \ref{amin} 
we depict the dependences of this
on the temperature and the plasma
parameter $\alpha$.
We see that both the temperature and the plasma parameter 
affect the acceleration rate significantly. 
%The correspondent value of $\mu_{*}$ is shown on Figure \ref{mumin}. 
As expected the value of $f_{turb}^{*}$ goes to zero when either
temperature goes to zero (cold plasma case) or when we deal with a low
density high magnetic field plasma ($\alpha \rightarrow 0$).  
This threshold 
also depends on the electron's energy. This dependence is shown on
Figure \ref{rmin} for several values of the plasma parameter and
temperature. We also note that electrons with larger values of $\mu$
(smaller pitch angles) are accelerated by the slower decaying, longer
wavelength waves. This effect is demonstrated on Figures \ref{mumin},
\ref{muminT}, and \ref{muminR} where we plot 
the value of the pitch angle cosine
$\mu_{*}$, at the minimum of the curves in Figures \ref{ratA} and
\ref{ratT}, versus $\alpha$ (for different temperatures), versus
temperature (for different $\alpha$) and versus $\beta/\beta_{th}$ for different values of
$\alpha$ and $T$. Note that for the majority of cases $\mu_* >
0.75$ which indicates that the pitch angle of the accelerated particles
will be somewhat anisotropic being beamed along the field lines.
This is especially true at lower energies when the scattering time is
much larger than the acceleration time or the wave decay time.

%The dependence of the $f_{turb}$ on the spectral index comes from the term
%$k^{q}$ in the evaluation of the acceleration time as well as from the
%normalization coefficient $(q-1) k_{min}^q$ in the definition of
%$\tau_p^{-1}$.
% Thus the value of
%$f_{turb}$ will increase with increasing $q$ as can be seen on Figure
%\ref{ratA}, hence decreasing the
%spatial fraction of the accelerated electrons at given turbulence
%level $\phi f_{turb}^{tot}$. But since $\phi \propto (q-1) k_{min}^q
%\simeq (q-1)$ this smaller fraction will correspond to the smaller
%value of $f_{turb}^{tot}$ hence affecting very little the dependence
%of the fraction of the accelerated electrons on the total level of plasma
%turbulence. 
%Here and in what follows we assume the spectral index $q=5/3$.

%Figure \ref{muminR} shows that the correspondent value of
%$\mu_{*}$ has a minimum value for particles with energy close to the
%thermal energy ($\beta \simeq \beta_{th}$). 
%In order to estimate the effectiveness of the acceleration at a given
%turbulence level we compute the fraction of the plasma particles which
%have the acceleration rate greater than the wave decay rate. 

\begin{figure}[h]
\leavevmode\centering
\psboxto(\textwidth;0pt){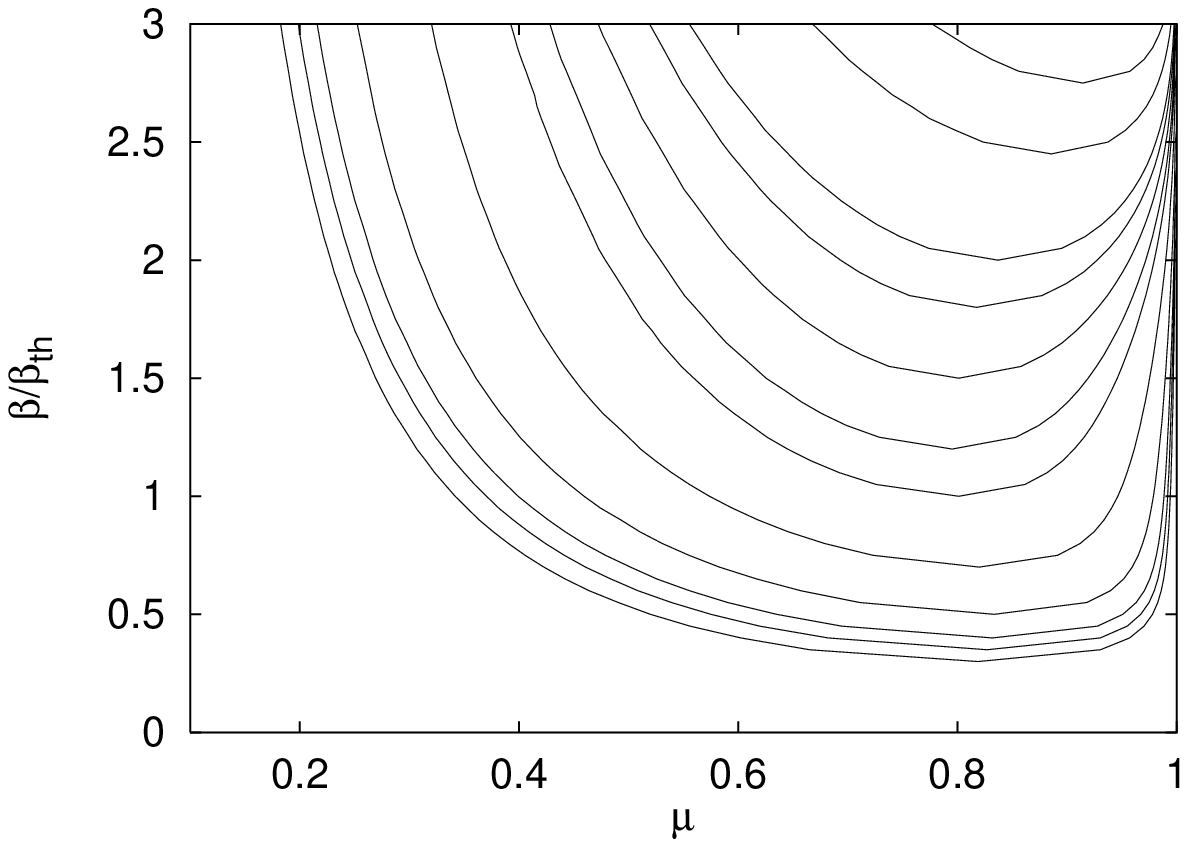}
\caption{The dependence of the minimum of turbulent ratio $f_{turb}$ 
on electron's velocity and pitch angle cosine
in plasma with $\alpha = 1$ and $T=0.1$ keV. The turbulent energy
fraction decreases from bottom to top of the plot with values of
$f_{turb}$: 0.016, 0.012, 0.008, 0.005, 0.002, 0.001, 0.00085,
0.0006, 0.0004, 0.0003, 0.00016, 0.00007, respectively.}
\label{mucont}
\end{figure}

Because of the complicated dependence of the pitch angle range
$(\mu_2 - \mu_1)$ on plasma parameters and on the electron
velocity $\beta$ the determination of the exact number of electrons
that can be accelerated is not straightforward.  We estimate this
fraction as follows.  On Figure \ref{mucont} we show a contour plot of
constant $f_{turb}$ in the velocity - pitch angle cosine plane. For
$f_{turb}$ equal to or grater than the value for each curve the
electrons with values of $\beta/\beta_{th}$ and $\mu$ above 
this curve will be
accelerated. We can then calculate the fraction of accelerated
electrons as a function of the turbulence level and other plasma parameters.

Assuming an isotropic Maxwellian distribution of the background
electrons the fraction of the accelerated electrons is given by
\beq\label{fraction} F(f_{turb}) =
\frac{4}{\sqrt{\pi}}\int_{r_{min}(f_{turb})}^{\infty} (\mu_2(r,
f_{turb}) - \mu_1(r, f_{turb})) r^2 e^{-r^2} d r, \eeq 
where $r=\beta/\beta_{th}$ and $\mu_1$,
$\mu_2$ and $r_{min}$ are obtained from the contour plot in Figure
\ref{mucont}. 
This fraction can be represented as a sum of two parts:
$F(f_{turb}) = F_{r < 3} + F_{r > 3}$. Note that for $r_{min} \leq 2$
the second term is less than 1\% of the total value. We evaluate
the first term numerically and use the analytical solution
(\ref{anrat}) found in \S 3 to estimate the second term.

\begin{figure}[h]
\leavevmode\centering
\psboxto(\textwidth;0pt){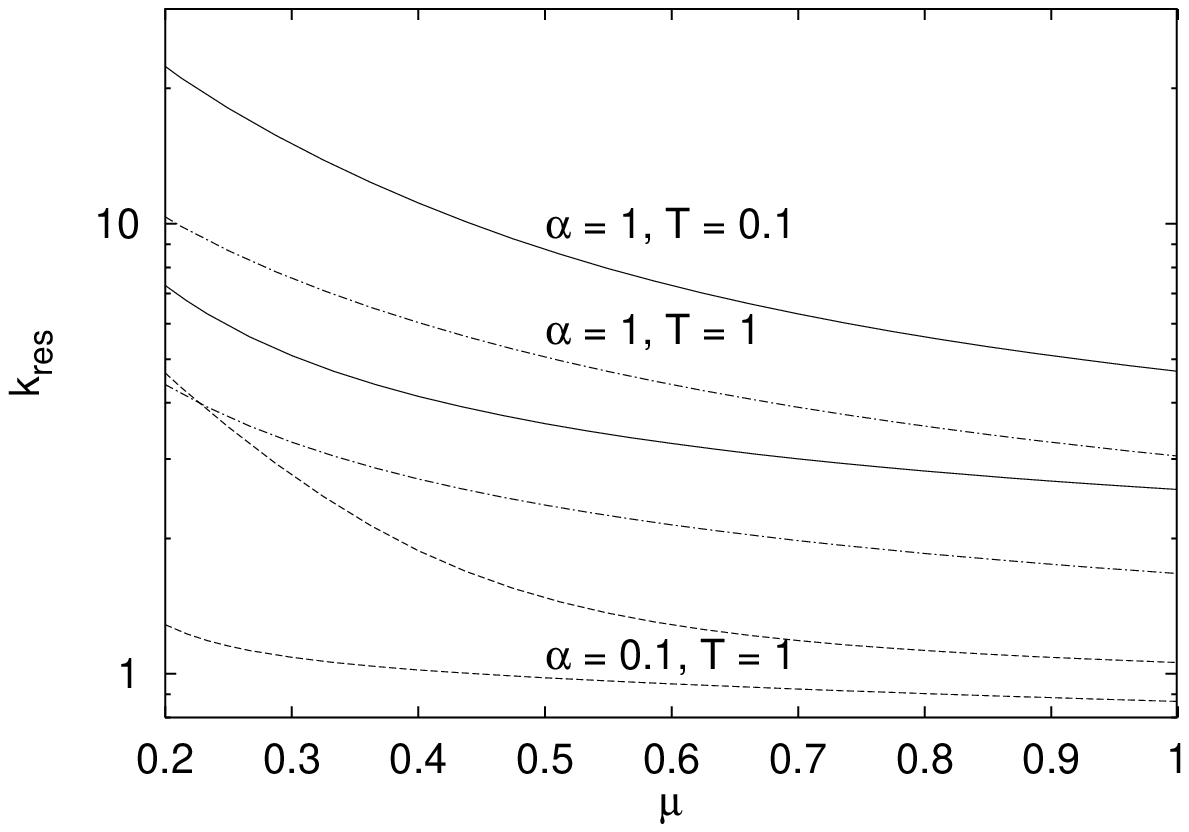}
\caption{The dependence of the resonant wavevector on cosine of the
  electron's pitch angle. Solid lines
  are for a plasma with $\alpha =1$ and $T=0.1$ keV, dashed
  lines for  $\alpha =0.1$ and $T=1$ keV, and dashed-dotted
  lines for  $\alpha =1$ and $T=1$ keV. In each case the upper curve is
for electrons with $\beta=\beta_{th}$
  and lower curve for $\beta=3 \beta_{th}$.}
\label{kres}
\end{figure}

\begin{figure}[h]
\leavevmode\centering
\psboxto(\textwidth;0pt){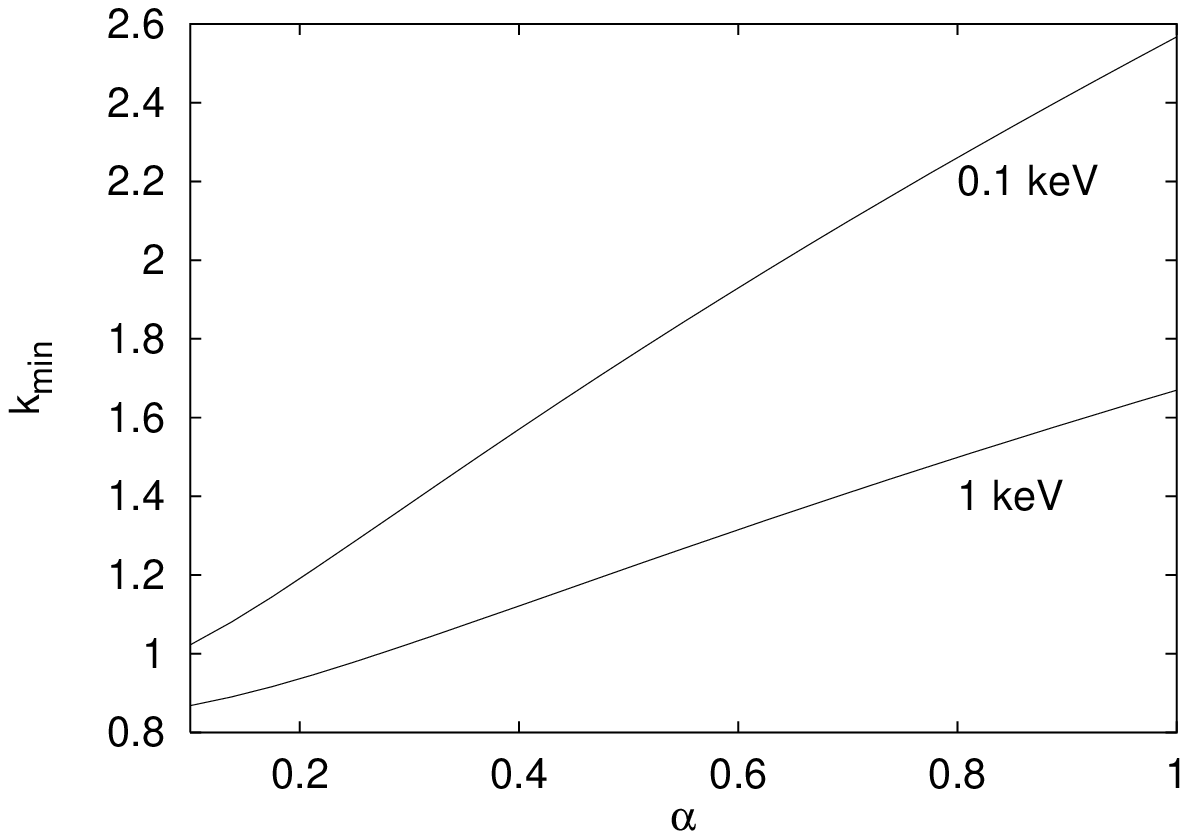}
\caption{The dependence on plasma parameter $\alpha$ 
of the minimum resonant wavevector 
of plasma waves which interact with electrons of 
velocities up to $3 \beta_{th}$  for two
different temperatures.}
\label{kmin}
\end{figure}

Since the main contribution to the fraction of the accelerated
electrons comes from the thermal electrons with $\beta \leq 3
\beta_{th}$ we can estimate the minimum wave number $k_{min}$ of waves
in resonance with these electrons. This will allow us to determine the
required turbulence $f_{turb}(k > k_{min})$ corrected for the
$k_{min}^{q-1}$ term discussed earlier. 
On Figure \ref{kres} we show the
dependence of the resonant wave vector $k_{res}$ on the cosine of
electron's pitch-angle for $\beta =\beta_{th}$ and $\beta =
3\beta_{th}$ and different values of $\alpha$ and $T$. From
this plot we see that in most cases the value of the resonant wave
vector is greater than or about unity.  Note that $k_{res}(\mu)$ decreases
monotonicly with $\mu$ and we use the minimum value at $\mu = 1$ which
gives the highest demand on the required amount of turbulence. We show the
dependence of $k_{min}$ on plasma parameter for two different
temperatures on Figure \ref{kmin}. The value of $k_{min}$ increases
with $\alpha$ and decreases with $T$ but remains of order of unity in
the considered range of plasma conditions.

\begin{figure}[h]
\leavevmode\centering
\psboxto(\textwidth;0pt){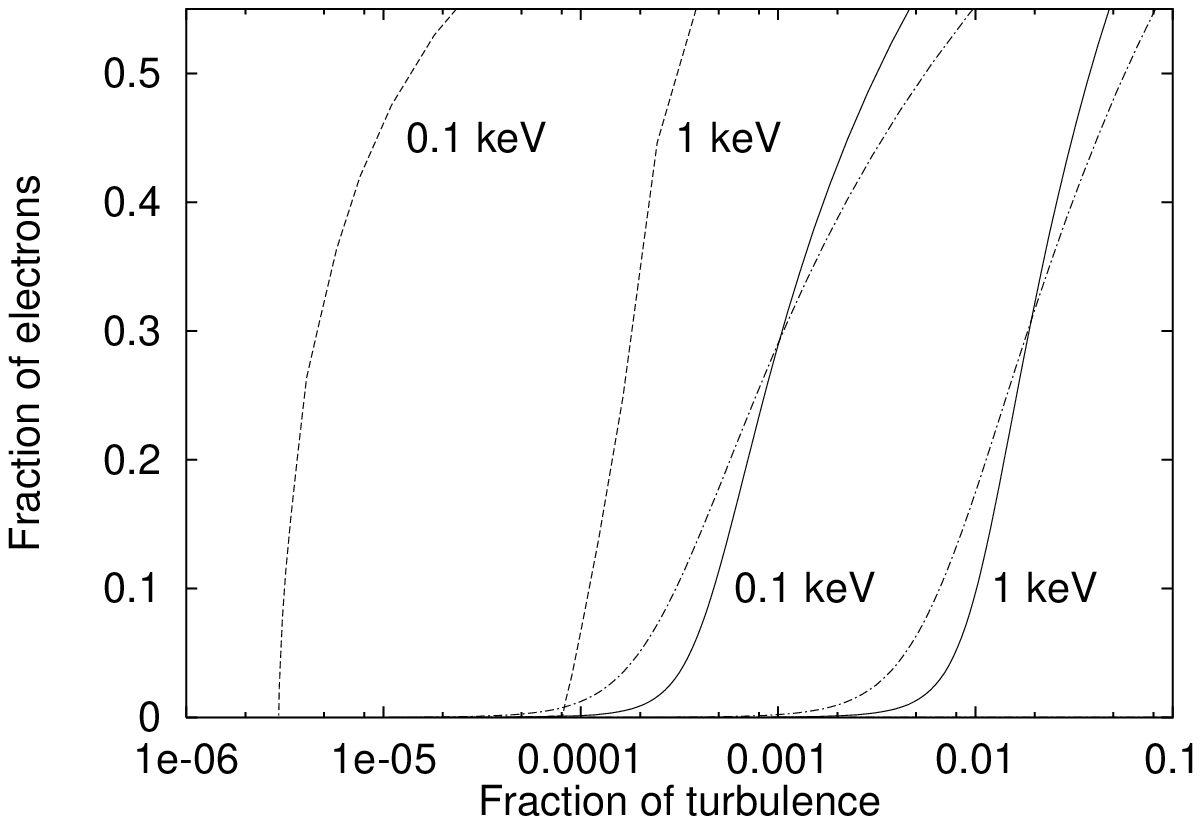}
\caption{The fraction of isotropicly distributed thermal electrons  
  involved in the acceleration process as a function of the total fraction
  of turbulence level $f_{turb^{tot}}$, assuming a power law up to
  $k_{min}^{th}$ taken from Figure \protect\ref{kmin}, for two
  temperatures, $0.1$ and $1$ keV.  Dashed lines are for $\alpha =
  0.1$, $q=5/3$, solid lines for $\alpha=1$, $q=5/3$, dashed-dotted
  lines for $\alpha = 1$, $q=3$. }
\label{totFrac}
\end{figure}

On Figure \ref{totFrac} we plot $F(f_{turb}^{tot})$ for different values of
$\alpha$, $T$ and spectral index $q$ and using the
values of $k_{min}$ from Figure \ref{kmin}.  As can be seen from this
plot the fraction of electrons which can be accelerated in a 
time shorter than it takes for the turbulent waves to decay increases rapidly
with increasing of the level of turbulence  (as a fraction of the total
magnetic energy density). As we noticed earlier, less turbulence is 
required for acceleration of $10$ \% of the background electrons with
higher value of the spectral index $q$. As expected from the analytic result
obtained in \S 3, the turbulent level required for the acceleration of
some given fraction of particles strongly depends on plasma parameter
$\alpha$ and temperature.

\section{ESTIMATION OF THE TOTAL LEVEL OF PLASMA TURBULENCE}

In the previous section we have considered acceleration of low energy
($\beta \leq 3 \beta_{th}$) electrons which represent a majority of
the electrons in a Maxwellian distribution. For most purposes we need
further acceleration to much higher energies. This process will
require presence of more turbulence at lower $k$ values or longer
wavelengths.  We now compare the level of turbulence required for the
initial acceleration, when thermal effects are important, with that
required for the acceleration beyond $3\beta_{th}$ when the thermal
effects can be ignored and the cold plasma relations obtained in PP1
are applicable.  Using the results presented in Figure \ref{totFrac}
we calculate the value for the level of
turbulence with $k \ge k_{min}^{th}$, 
needed for acceleration of 10\% of thermal electrons.
Table 1 summarizes these results.

%In this section we compare the results obtained for the
%non-relativistic electrons with those for the relativistic electrons
%(see PP1). In both cases we assumed that a power law wave spectrum is
%maintained down to some minimal wave vector $k_{min}$. Since in the
%estimation of the fraction of the accelerated electrons
%(\ref{fraction}) we considered only electrons with $\beta < 3
%\beta_{th}$ we can estimate $k_{min}^{ec} = k(\mu =1, \beta = 3
%\beta_{th})$ for given $\alpha$ and $T$.  Note
%that $k(\mu)$ is a monotonicly decreasing function and the value of
%$k_{min}$ increases with $\alpha$ and decreases with temperature and
%for the conditions under consideration it is of order of unity.  Using
%the definition of $\tau_p$ in equation (\ref{taup}) we can estimate
%the fraction of the total turbulence $f_{turb}^{tot}$ due to the
%waves with $k > k_{min}^{ec}$ needed to accelerate a given fraction of
%the background electrons (we use the value 10\%) for different plasma
%conditions. The results of such estimation are given in Table 1.

\begin{table}[htbp]
  \begin{center}
    \leavevmode
    \begin{tabular}[c]{|l|c|c|c|}
      % use r to right-align, l to left align, c to center.  See
      % attached info page!
      \hline
      $T$, keV\ &$\alpha = 0.1, q = 5/3$&$\alpha = 1, q = 5/3$&$\alpha = 1, q = 3$\\
      \hline
      $0.1$&$3\ 10^{-6}$&$5\ 10^{-4}$&$3\ 10^{-4}$\\      
      \hline
      $1.0$&$10^{-4}$&$10^{-2}$&$6\ 10^{-3}$\\
      \hline
    \end{tabular}
  \end{center}
  \caption{Turbulent energy density as a fraction of the
    magnetic energy density needed for acceleration of 10\% of the background
    electrons.
 }
\end{table}

%The acceleration time also depends on the
%electron gyrofrequency   Here and in what
%follows we assume the value of magnetic field $B$ equal to 100 Gs and
%1000 Gs for 

The amount of turbulence required for further acceleration to higher
energies depends on several factors. The first one is
the required acceleration
time scale $\tau_a$. For shorter times we need proportionally higher
values of turbulence. The second factor is the electron gyrofrequency 
$\Omega_e = e B /m_e c $. The third factor is the plasma parameter $\alpha$. In
what follows we assume $\tau_a \Omega_e \simeq 10^{10}$ and $10^{11}$ 
for $\alpha = 1$ and $\alpha = 0.1$, respectively. This
corresponds to the acceleration within a second in plasmas with density
$n = 10^{9} {\rm cm}^{-3}$ and uniform magnetic fields of $100$ and $1000$
Gs, respectively.
Another factor which affects $f_{turb}$ is the shape of its
spectrum. For a power law distribution (as assumed in PP1)
this depends on the index $q$ and the value of $k_{min}$. 
Using the results from PP1  we show the required values of the
turbulence with $k_{min}^{rel} < k < k_{min}^{th} \simeq 1$ for several
plasma parameters in Table 2. For each energy we use the
correspondent value of $k_{min}^{rel}$, the minimum wave number
of the waves which are in resonance with the relativistic electron. 

%Using the similar approach one can estimate the fraction of the total
%turbulence carried by the waves with $k_1 < k < k_{min}^{ec}$, needed for
%the acceleration of high-energy electrons. The minimum wavenumber
%$k_1$ for Alfv\'en waves is usually considered to be of
%order of $10^{-5}$ (Miller \& Ramaty, 1992). We shown in \S 3 that for
%electrons with $\beta \gg \beta_{th}$ the results obtained for the
%cold plasma case are still valid. Assuming the average
%acceleration time for the relativistic electrons to be of order of one
%second and using the results represented on Figure 13 of PP1 we
%estimate the needed level of turbulence for different plasma
%conditions and energies of electron. Since $\tau_p$ also depends on
%the electron's gyrofrequency ( and thus, on the magnetic field) we
%assume $B=100$ Gs in the plasma with $\alpha = 1$ and $B=1000$ Gs in
%the plasma with $\alpha=0.1$. We summarize these results in 
%Table 2.

\begin{table}[htbp]
  \begin{center}
    \leavevmode
    \begin{tabular}[c]{|l|c|c|c|c|c|}
      % use r to right-align, l to left align, c to center.  See
      % attached info page!
      \hline
      $E$, MeV\ &$k_{min}$&$\alpha = 0.1, q=5/3$&$\alpha = 0.1,q=3$&$\alpha = 1,q=5/3$&$\alpha = 1,q=3$\\
      \hline
      $10^4$&$5\  10^{-5}$&$1.5\ 10^{-5}$&$3\ 10^{-5}$&$10^{-2}$&$1.6\ 10^{-2}$\\      
      \hline
      $10^3$&$5\  10^{-4}$&$10^{-6}$&$3\ 10^{-6}$&$10^{-3}$&$1.6\ 10^{-3}$\\      
      \hline
      $10^2$&$5\  10^{-3}$&$2.5\ 10^{-8}$&$3\ 10^{-8}$&$5\ 10^{-5}$&$1.6\ 10^{-4}$\\      
      \hline
    \end{tabular}
  \end{center}
  \caption{Turbulent energy density as a fraction of the 
    magnetic energy density needed for acceleration of high-energy electrons.
 }
\end{table}

From comparison of the values in the two tables one can conclude that
in order to have a successful acceleration of at least 10\% of the 
background electrons with the subsequent acceleration to GeV
energies on the
time scale of $\tau_a = (10^{10} - 10^{11})/\Omega_e$, 
the turbulent energy carried by the waves with
$k < k_{min}^{th}$ should be roughly the same or less than the energy due to the turbulent
waves with $k > k_{min}^{th}$. This suggests that the turbulent spectrum could
flatten at lower $k$ values having a smaller value of spectral index $q$.
Assuming that one half
of the turbulent energy is due to the waves with high wave numbers we
can give the very conservative estimation of the total level of the
plasma turbulence $f_{turb}^{tot}$ by doubling the numbers given in Table 1.
The very steep behavior of the curves on Figure \ref{totFrac}
suggests that a small increase in the turbulence level will lead to
a significant increase in the fraction of the accelerated background
electrons.

%In our previous paper we have shown that the acceleration process
%strongly depends on the value of the plasma parameter and a new
%transport equation was obtained for low-energetic electrons in
%plasmas with $\alpha\simeq 0.1-0.3$. Since we want to determine the
%conditions under which the cold plasma results can be still applicable
%we will concentrate on this range of plasma parameter. Figure
%\ref{tmin} shows that in the wide range of plasma temperature the
%turbulent threshold is less than or of order of $10^{-4}$.

\section{SUMMARY}

This paper extends our earlier work (PP1) on the acceleration of
low-energy electrons by plasma turbulence to include the effects of
the finite temperature of the plasma.  We investigate the possibility of
acceleration of the low energy background (thermal) electrons by this
process. We use the well known formalism developed over the years and
specifically the formalism proposed by Schlickeiser, 1989 and Dung \&
Petrosian, 1994.  As in our previous paper we consider interaction of
electrons with plasma waves propagating along the magnetic field
lines.  In a finite temperature plasma the high wave number, short
wavelength waves needed for the acceleration of low energy electrons are
damped. The rate of the damping is given by the imaginary part of the
resonant wave frequency $\omega_i$.  This does not preclude acceleration by
waves of thermal particles. Whether or not thermal electrons will be
accelerated depends on whether or not the damping rate of the waves,
which can accelerate an electron with velocity $\beta$ and pitch angle
$\mu$, is smaller or larger than the acceleration rate of such an
electron.

In the first case the conditions for acceleration will be similar to
those found for cold plasmas. In the second case acceleration will be
possible if turbulence is generated at a rate equal to or faster than
the damping rate. This, of course, increases the energy demand for the
acceleration process. In this paper we determine the plasma conditions
in which the acceleration rate is faster than the damping rate. Since
the acceleration rate depends on the level of turbulence present in
the plasma this gives us a lower limit on the level of
the required turbulence as a function of other plasma parameters.
%In this case the resonant interaction of electrons with
%transverse plasma branch can be described by equations similar to
%those obtained for the cold plasma if the sufficient level of
%turbulence is present.
%we determine the required
%energy density of waves as a fraction of the magnetic energy density
%so that a substantial fraction of the background plasma electrons can
%be accelerated.  As it frequently the case in plasma physics the

We determine this level from comparison of the acceleration rate of
the particles with the decay rate of the waves they interact with.
The dispersion relation for finite temperature plasma is complex so
that a 
simple analytic solution to the problem is not always possible.
In the case of electrons for which $\mu
\beta/\beta_{th} \gg 1$ we find an asymptotic analytic solution for
the ratio of acceleration to decay times, $\tau_a/\tau_{dec} = \tau_a
\omega_i$. For electrons with $\beta < 3 \beta_{th}$ we find the exact
solution numerically. Since $\tau_a^{-1}$ is proportional to the fraction of
plasma turbulence $f_{tot}={\cal E}_{turb}/(B_0^2/8\pi)$, the
condition $\tau_{a} \omega_i < 1$ sets a lower limit on the amount of
turbulent energy that is needed for acceleration of plasma electrons
with energies of order of the thermal energy.
%We approximate the wave spectrum as that given initially up
%to a critical $k_{cr}$ (at which the damping and acceleration rates
%are equal) beyond which the wave spectrum cuts off abruptly. The
%energy and pitch angle range of electrons which are in resonance with
%waves of $k < k_{cr}$ then determine the fraction of electrons that are
%accelerated efficiently.
%Using the relation  we find a function 
%$f_{turb}(\mu,r,\alpha,\beta,q)$ which is the minimum of 
%plasma turbulence required for acceleration. 
We investigate the dependence of this minimum value of the turbulence,
$f_{turb}^{min}$, on plasma parameters and electron's pitch angle and
velocity. We show that the required level of turbulence increases
with the plasma parameter
$\alpha = \omega_{pe}/\Omega_e$ (the ratio of electron plasma
frequency to electron gyrofrequency) and the temperature, and it
is almost independent of the spectral index $q$ and the pitch angle 
for a wide range of pitch angles.

Assuming an isotropic Maxwellian  distribution of the background
electrons we estimate the approximate fraction of these electrons that
can be accelerated for a given turbulence level $f_{turb}(k >
k_{min}^{th}(\alpha, T))$. We show that the main contribution to
this fraction comes from electrons with $\beta \leq 3 \beta_{th}$
which allows us to determine the minimum wave number $k_{min}^{th}$ of
the turbulent waves in resonance with these electrons. The value of
$k_{min}^{th}$ increases with $\alpha$ and decreases with $T$ and is
of order of unity.  The fraction of accelerated electrons increases
with decreasing $\alpha$ and temperature and a small increase in the
turbulence level beyond the minimum level can lead to a significant increase in the fraction of
the accelerated background electrons.

We compare the level of turbulence required for the initial
acceleration of electrons with velocities of order of thermal
velocity, when thermal effects are important, with that required for
further acceleration to higher energies, when the thermal effects can
be ignored. We conclude that in order to have a successful
acceleration of at least 10\% of the background electrons with the
later acceleration of electrons to GeV energies on the time scale of
$10^{10}\Omega_e^{-1}$ to $10^{11}\Omega_e^{-1}$, the turbulent energy
carried by the waves with $k < k_{min}^{th}$ should be roughly the
same or higher than the energy due to the turbulent waves with $k >
k_{min}^{th}$.  Thus a very conservative estimate of the total level
of the plasma turbulence $f_{turb}^{tot}$ is obtained by assuming that
the turbulent energy is equally distributed between the waves with
high and low wave numbers.  This estimated value does not exceed 2\%
in the worst case scenario.

This work was supported by NSF grant ATM 93-11888 and NASA grant NAGW 1976.

\end{document}